\documentclass[showpacs, pra,onecolumn,preprintnumbers ,amsmath, amssymb, superscriptaddress, aps]{revtex4-2}
\usepackage{color}
\usepackage{amsmath,amssymb}
\usepackage{pifont}
\usepackage{amssymb}  
\usepackage{bbold}
\usepackage{float}
\usepackage{subfloat}

\usepackage[caption=false]{subfig}
\usepackage{tikz}
\usepackage{makecell}
\usepackage{subfig}
\usepackage{pifont}   
\usepackage{graphicx} 
\graphicspath{{Figures/}}
\usepackage{dcolumn}  
\usepackage{bm}       
\usepackage{multirow} 
\usepackage{placeins}
\usepackage[colorlinks]{hyperref}
\usepackage{mathtools}
\usepackage{appendix}

\def \be{\begin{align}}
	\def \ee{\end{align}}
\def \bea{\begin{eqnarray}}
	\def \eea{\end{eqnarray}}

\begin{document}
	
	\title{Electrons trapped in graphene magnetic quantum dots with mass term}
		
	\author{Mohammed El Azar}
	\affiliation{ Laboratory of Theoretical Physics, Faculty of Sciences, Choua\"ib Doukkali University, PO Box 20, 24000 El Jadida, Morocco}
	\author{Ahmed Bouhlal}
	\affiliation{ Laboratory of Theoretical Physics, Faculty of Sciences, Choua\"ib Doukkali University, PO Box 20, 24000 El Jadida, Morocco}
	\author{Ahmed Jellal}
	\affiliation{ Laboratory of Theoretical Physics, Faculty of Sciences, Choua\"ib Doukkali University, PO Box 20, 24000 El Jadida, Morocco}
	\affiliation{
		Canadian Quantum  Research Center,
		204-3002 32 Ave Vernon, BC V1T 2L7,  Canada}

	\begin{abstract}
	
		{Owing to the Klein tunneling phenomenon, the permanent confinement or localization of electrons within a graphene quantum dot is unattainable. Nonetheless, a constant magnetic field can transiently ensnare an electron within the quantum dot, giving rise to what are known as "quasi-bound states" characterized by finite lifetimes. To prolong the retention of electrons within the quantum dot, we introduce a mass term into the Hamiltonian, thereby inducing an energy gap. We resolve the Dirac equation to ascertain the eigenspinors, and by ensuring their continuity at the boundaries, we investigate the scattering behavior. Our findings indicate that the presence of an energy gap can extend the lifetimes of these quasi-bound states within the quantum dot. In particular, we demonstrate that even in the absence of a magnetic field, the scattering efficiency attains significant levels when the energy gap gets closed to the incident energy of an electron traversing the quantum dot. It is found that an augmentation in the electron density within the quantum dot results in an enhancement of the electron-trapping time.}

	\end{abstract}
	\pacs{81.05.ue; 81.07.Ta; 73.22.Pr\\
	{\sc Keywords}: Graphene, circular quantum dot, magnetic field, energy gap, scattering phenomenon}

	\maketitle
	
	\section{Introduction}
	
	A two-dimensional semi-metal called graphene is characterized by a linear band structure that is quite close to the Fermi level. It has a honeycomb lattice structure comprised entirely of carbon atoms. Researchers are very driven to enhance the conclusions established theoretically or experimentally in this context (see  \cite{novoselov2004electric, novoselov2005two}) because of the very specific electrical characteristics of graphene. The anomalous quantum Hall effect \cite{gusynin2005unconventional, ostahie2015electrical}, Klein tunneling \cite{katsnelson2006}, high electrical conductivity, and extremely high electron mobility \cite{geim2007rise, geim2009graphene} are only a few of the extraordinary electronic characteristics of graphene.
	Several scientific studies have demonstrated graphene's interest in fundamental physics for more technological applications in a variety of fields (e.g., \cite{neto2009, abergel2010, kotov2012}). {In-depth studies of the physics of materials like graphene are usually based on their interactions with external fields. As a result, they  showed} that it can offer a perfect framework to study fundamental physics and interpret physical effects and phenomena such as Landau level quantization \cite{guinea2006electronic, lukose2007novel, yin2015landau}, Aharonov-Bohm effect \cite{recher2007aharonov, heinl2013interplay, zebrowski2018aharonov}. Then graphene is one of the strongest materials tested so far, possessing {remarkable  qualities} that serve to use graphene in technological applications such as integrated circuits, light sensing devices, and microelectronic devices \cite{konstantatos2012hybrid, dani2012intraband, gruber2016ultrafast}.

{It is known that because of Klein tunneling, electrons cannot be localized in a small and constrained region of graphene via an applied electrostatic gate.} The main objective of graphene-based electronics is to confine {electrons, which generate} quantum dots \cite{silvestrov2007quantum, de2007magnetic, chen2007fock, pedersen2008graphene}. Since graphene does not have a band gap, {it does not have} any traditional quantum {dots that} can localize electrons in areas of finite dimensions. Utilizing the quantum dot for relativistic electrons, which behave like massless Dirac fermions, {will generate potential applications of graphene in electronic uses} \cite{novoselov2005two, katsnelson2006}. {It was shown that the}  normal incidence issue, which is the cause of perfect transmission (Klein paradox), is resolved by using zero-dimensional circular quantum dots \cite{chakraborty1990comments}. {Also, the spin-orbit interaction is used to modify the electrical and transport properties} of tiny graphene-like flakes (graphene-like quantum dots) \cite{PhysRevB.92.045427}. {Additionally, theoretical studies} of transport across weakly coupled external ferromagnetic leads in graphene quantum dots {(GQDs) have been achieved}
		\cite{PhysRevB.85.205306}. {By confining Dirac fermions in graphene, different applications of quantum dots in electronics are made possible}, including solar panels, lasers \cite{fafard1996red}, photo-detectors \cite{liu2001quantum}, quantum information processing, and quantum computers \cite{loss1998quantum}.

	{We recall that the localization of GQDs} is prevented by Klein tunneling when an electron strikes the dot with normal incidence \cite{katsnelson2006}. 
	The trapping of electrons in quantum dots is currently a very interesting research topic in the field of condensed matter physics. It began decades ago and continues to receive a lot of attention from both a fundamental and an application standpoint. The electron trapping problem was first studied in a one-dimensional wire in the absence of a magnetic field \cite{silvestrov2007quantum, trauzettel2007spin, brey2006electronic}, and then for a quantum dot with smooth \cite{chen2007fock} and sharp \cite{matulis2008quasibound} boundary states. The trapping potential of a relativistic electron in graphene  {depends} on several factors, in particular the transverse momentum \cite{silvestrov2007quantum, chen2007fock, cheianov2006selective}. {For} a quantum dot, the trapping potential is closely related to the  {angular momentum of the electron} and becomes significantly more intense {with its increase}.   
	Trapping an electron in a {GQD} is thus best accomplished when the {confining} potential is smooth and the electronic states have a large angular momentum. It is shown that Klein tunneling is significantly reduced in the resonant regime, and the lifetime increases with increasing magnetic field \cite{pena2022electron}.

{Creating an energy} gap between the two valence and conduction bands in {the energy spectrum} of  graphene  can be {realized by} using a variety of experimental procedures \cite{abergel2010}. {As a consequence of the symmetry breaking of graphene sublattices, the highest value of the energy gap is around} 260 meV \cite{zhou2007substrate}. {Note in passing that the value of the energy gap differs from one experiment to another and therefore depends on testing techniques.} {Furthermore, by manipulating the interface structure}  between graphene and ruthenium (Ru), {it is shown that there are alternative experimental ways for opening gaps} \cite{enderlein2010formation}. {Silicon carbide (SiC) is another example of a substrate that has been used to generate an energy gap in graphene} \cite{zhou2007substrate}. {Also}, different band gaps are created {by} depositing a graphene layer on other substrates \cite{rusponi2010highly, kim2018accurate}. {We close this part by mentioning that in gapped graphene, the density of states (DOS) was examined for electrostatic confinement and magnetic flux \cite{BOUHLAL2021114335}. It was found that DOS exhibited different oscillatory behaviors according to the values taken by the magnetic flux. Additionally, it was discovered that the energy gap can influence not only the position but also the amplitude and width of these resonances.}

{Motivated by the previously mentioned findings, particularly those presented in \cite{pena2022electron}, our investigation delves into the interaction of incident electrons with a circular GQD featuring an energy gap while being exposed to a constant magnetic field. Our focus centers on examining key aspects, including the scattering efficiency denoted as $Q$, the probability density represented by $\rho$, and the electron lifetime denoted as $\tau$.~These parameters serve as crucial indicators to illustrate the impact of the energy gap on the scattering phenomena within our system. We initiate our study by employing the Dirac equation to analytically derive solutions for the energy spectrum. Subsequently, we employ continuity conditions at the interface to compute the corresponding scattering efficiency both within and outside the quantum dot. Our findings reveal that when the energy gap approaches the incident energy of an electron traversing the GQD, the efficiency remains remarkably high, even in the absence of a magnetic field.}

	The present paper is organized as follows. In Sec. \ref{theory}, we establish a theoretical model describing our system and determine the solution of the energy spectrum. After matching eigenspinors at the interface, we explicitly determine the quantities characterizing the scattering phenomenon in Sec. \ref{SSPP}. In Sec. \ref{res}, we numerically analyze our finding under various conditions of the physical parameters, {where} the scattering efficiency $Q$, probability density $\rho$, and lifetime $\tau$ are {three} examples. Finally, we conclude our results.

	\section{Mathematical background}\label{theory}
\begin{figure}[h]
		\centering 
		\includegraphics[scale=0.4]{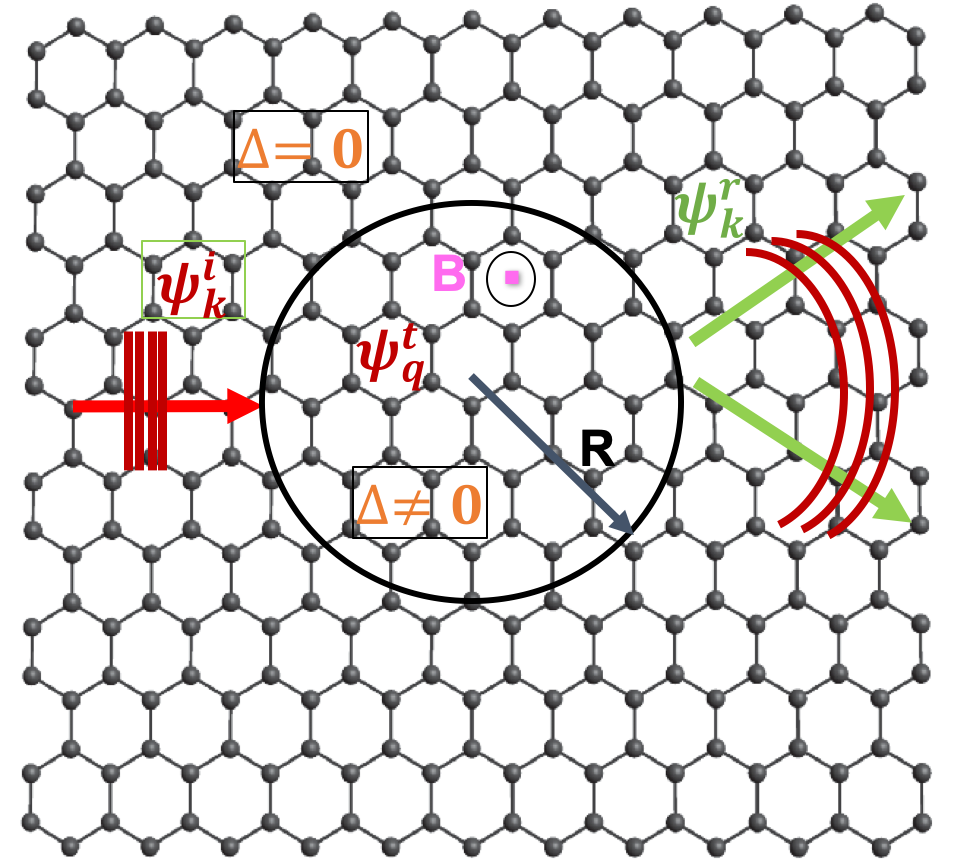}	
		\caption{(color online) A graphene quantum dot of radius $R$ is placed in the horizontal plane $xy$, with the magnetic field $B$ oriented perpendicular to the dot plane in the direction $z$. The incident electron is represented by a plane wave $\Phi_k^i$ with energy $E=\hbar v_F k$. When an electron approaches a quantum dot, it is either a reflected wave $\Phi_k^r$ or a transmitted wave $\Phi_q^t$. 
		}\label{figsystem}
	\end{figure}

{Let us consider a circular graphene quantum dot (GQD) with an energy gap and subjected to a constant magnetic field that} is made of two regions,  as depicted in Fig. \ref{figsystem}.   
\noindent 	
	Our system can be described by a single valley Hamiltonian as follows
	\begin{equation}\label{Hamilt}
		H =v_F \vec \sigma \cdot (\vec p +e \vec A)+\Delta\sigma_z
	\end{equation}
	where $v_F = 10^6$ ms$^{-1}$  is the Fermi velocity, {$\vec \sigma=(\sigma_x ,\sigma_y)$ and  $\sigma_z$} are  Pauli matrices, {$(-e)$ is the electron charge, and  $2\Delta$} is the energy gap resulting from the mass term. It convenient to select  the vector potential {$\vec A = \frac{B}{2}(-y,x)$} in symmetric gauge. Because of the spherical symmetry, we can write the Hamiltonian {in polar coordinates}
	\begin{equation}\label{ham2}
		H = \begin{pmatrix} \Delta & -i\hbar v_F  e^{-i\theta}\left(\partial_r - \frac{i}{r}\partial_\theta - \frac{e Br}{2\hbar}\right)\\
			-i\hbar v_F  e^{i\theta}\left(\partial_r + \frac{i}{r}\partial_\theta + \frac{e Br}{2\hbar}	\right)&-\Delta \\
		\end{pmatrix}	
	\end{equation}	
	{where we have used }
	\begin{equation}
		\sigma_r = \begin{pmatrix} 0 &\ e^{-i\theta}\\\\ \ e^{i\theta} &0 \\
		\end{pmatrix}, \quad 
		\sigma_\theta = \
		\begin{pmatrix} 0 &\ -ie^{-i\theta}\\\\ \ ie^{i\theta} &0 \\
		\end{pmatrix}.
	\end{equation}

	Given that the total angular momentum operator $J _z= -i\hbar \partial_ \theta +\frac{\hbar}{2}\sigma_z$  commutes with the Hamiltonian \eqref{ham2}, i.e., $ [H, J_z] = 0 $, the eigenspinors can be separated as
	\begin{equation}\label{ansatz}
		\Phi(r,\theta)= \dbinom{\Phi^A(r) e^{im\theta}}{i\Phi^B(r)e^{i(m+1)\theta}}  
	\end{equation}
	where the integer values $m$ are  eigenvalue of  $J _z$. Now using the eigenvalue equation	
	$H \Phi(r,\theta)=E \Phi(r,\theta)$ to obtain 	
	\begin{subequations}\label{8}
		\begin{align}
			&			\left(\partial_r + \frac{r}{2 l_B^2} -\frac{m}{r}\right) \Phi^A(r) =-\frac{E+\Delta}{\hbar v_F} \Phi^B(r) \label{8a}\\
			&	\left( \partial_r+\frac{m+1}{r} -\frac{r}{2l_B^2}\right) \Phi^B(r) = \frac{E-\Delta}{\hbar v_F} \Phi^A(r) \label{8b}.
		\end{align}
	\end{subequations}	
	For instance,	 by injecting  \eqref{8a} into \eqref{8b}, we end up with  a second differential  equation for $\Phi^A(r)$	
	\begin{equation}\label{e:9}
		\left( \partial_r^2+\frac{1}{r}\partial_r  + \frac{m+1}{l_B^2}-\frac{r^2}{4 l_B^4}-\frac{m^2}{r^2}+q^2\right)\Phi^A(r) =0
	\end{equation}
	where we have set  $q=\frac{\sqrt{\vert E^2-\Delta^2\vert}}{\hbar v_F}$ and   $l_B= \sqrt{\frac{\hbar}{eB}}$ {is the magnetic length}.
	In order to solve \eqref{e:9}, we start by exploring the asymptotic limits that define the necessary physical behaviors depending on the value of $r$. {Indeed,} in the limit $r\rightarrow \infty$,  \eqref{e:9} can be approximated by
	\begin{equation}\label{e:10}
		\left( \partial_r^2+\frac{1}{r}\partial_r - \frac{r^2}{4 l_B^4}\right)  \Phi^A(r)=0
	\end{equation} 
{which is an equation of the Bessel type for the zero-order case and thus has the solution}
	\begin{equation}
		\Phi^A (r) =c_1 I_0\left( \frac{r^2}{4l_B^2}\right) +c_2 K_0\left( \frac{r^2}{4l_B^2}\right) 
	\end{equation}
	where $I_0(x) $ and $ K_0(x) $ denote the zero-order modified Bessel functions of the first and second kinds, respectively. We choose $c_1 = 0$ and $c_2 = 1$ to avoid the divergence of function $I_0(x) $ when $ x $ reaches infinity.  Now using the asymptotic behavior
	$ K_0(x) \underset{x\gg 1}{\sim} \frac{e^{-x}}{\sqrt{x}}$, to approximate  $\Phi^A(r)$ as
	\begin{equation}\label{e:13}
		\Phi^A (r) \sim 2 l_B\frac{e^{-\frac{r^2}{4 l_B^2}}}{r}.
	\end{equation}
	In the limit $r\rightarrow 0$,  \eqref{e:9} reduces to	
	\begin{equation}\label{e:14}
		\left(  \partial_r^2+\frac{1}{r}\partial_r  - \frac{m^2}{r^2} \right) \Phi^A(r)=0
	\end{equation}	
	which has the following  solution	
	\begin{equation}\label{e:15}
		\Phi^A(r) =\frac{c_3}{2}(r^m +r^{-m})+\frac{i c_4}{2}(r^m -r^{-m})
	\end{equation}	
	where  $c_3$ and $c_4$ must be chosen in such a way that the solution adheres to the physical constraints. {Then, we independently examine the positive and negative values of the angular momentum $m$}. Indeed, for $m\ge 0$,  $\sim r^{-m}$  must vanish,  {implying} that $c_4 = -ic_3$, and, again by convention, we put $c_3 = 1$. As for  $m< 0$,  $\sim r^m$  {has to vanish}, then we replace $c_4 = ic_3$ with $c_3 = 1$. Combining all to write the asymptotic  behavior of $\Phi^A(r)$ as
	\begin{subequations}\label{16}
		\begin{align}
			&\Phi^A(r) \sim r^m ,\qquad m\ge 0, \label{16a}\\
			&\Phi^A(r) \sim r^{-m},\qquad m<0 \label{16b} .
		\end{align}
	\end{subequations}
	Using the above analysis to write the solution of \eqref{e:9} as	
	\begin{equation}\label{e:19}
		\Phi^{A\pm} (r) =r^{\pm m }  \frac{e^{-r^2/4 l_B^2}}{r/2 l_B} \Xi_q^{A\pm}(r)
	\end{equation}	 
where the sign "+" stands for $ m\ge 0 $, while "-" stands for the opposite case, $ m<0 $. 
	Now, we perform the  variable change $ \eta=\frac{r^2}{2 l_B^2} $ and  use the transformation $ \Xi_q^{A\pm}(\eta)=\sqrt{\eta} \chi_q^{A\pm}(\eta)$ to write \eqref{e:19} as
	\begin{equation}\label{e:20}
		\Phi^{A\pm} (\eta) =l_B^{\pm m}  2^{\frac{1\pm m}{2}}  \eta^{\pm m/2 }  e^{-\eta/2} \chi_q^{A\pm}(\eta)
	\end{equation}
{and insert it} into \eqref{e:9} to end up with the Kummer-type differential equations	
	\begin{subequations}\label{21}
		\begin{align}
			&\eta \partial_\eta^2 \chi_q^{A+}(\eta) +\left( m+1-\eta\right)  \partial_\eta \chi_q^{A+}(\eta) +\frac{l_B^2 q^2}{2} \chi_q^{A+}(\eta)=0  \label{21a}\\
			&\eta \partial_\eta^2 \chi_q^{A-}(\eta) +\left( 1-m-\eta\right)  \partial_\eta \chi_q^{A-}(\eta)
			+\left( m+\frac{l_B^2 q^2}{2}\right)  \chi_q^{A-}(\eta)=0 \label{21b} 
		\end{align}
	\end{subequations}
which have the confluent hypergeometric functions as
solutions
	\begin{subequations}\label{18}
		\begin{align}
			&\chi_q^{A+}(\eta)=\prescript{}{1}{F}_1^{}\left(-\frac{l_B^2 q^2}{2},m+1,\eta\right) \label{18a}\\
			&\chi_q^{A-}(\eta)=\prescript{}{1}{F}_1^{}\left(-m-\frac{l_B^2 q^2}{2},1-m,\eta\right) \label{18b} .
		\end{align}
	\end{subequations}
	{Finally, the solutions to the second order differential equation \eqref{e:9} can be written as}	
	\begin{subequations}\label{22}
		\begin{align}
			&\Phi^{A+}(r)=r^{\vert m\vert} e^{-r^2/4 l_B^2}\prescript{}{1}{F}_1^{}\left(-\frac{l_B^2 q^2}{2},m+1,\frac{r^2}{2 l_B^2}\right) \label{22a}\\
			&\Phi^{A-}(r)=r^{\vert m\vert} e^{-r^2/4 l_B^2}\prescript{}{1}{F}_1^{}\left(-m-\frac{l_B^2 q^2}{2},1-m,\frac{r^2}{2 l_B^2}\right) \label{22b} 
		\end{align}
	\end{subequations}  
	{and the} other components of spinor \eqref{ansatz} can be obtained by {injecting} \eqref{22a} and \eqref{22b} into  \eqref{8a} {to get}	
	\begin{subequations}\label{24}
			\begin{align}
				&\Phi^{B+}(r)=\frac{q}{2(m+1)}r^{\vert m\vert+1} e^{-r^2/4 l_B^2}\prescript{}{1}{F}_1^{}\left(1-\frac{l_B^2 q^2}{2},m+2,\frac{r^2}{2 l_B^2}\right) \label{24a}\\
				&\Phi^{B-}(r)=\frac{1}{q}r^{\vert m\vert-1} e^{-r^2/4 l_B^2}\left[2m \prescript{}{1}{F}_1^{}\left(-m-\frac{l_B^2 q^2}{2},1-m,\frac{r^2}{2 l_B^2}\right)+ \frac{( 2m+l_B^2 q^2)r^2} {2( 1-m) l_B^2}\prescript{}{1}{F}_1^{}\left(1-m-\frac{l_B^2 q^2}{2},2-m,\frac{r^2}{2 l_B^2}\right)\right]  \label{24b}.
			\end{align}
	\end{subequations}
In the forthcoming analysis, we will see how the  above results can be used to study the scattering phenomenon in terms of the physical parameters, including incident energy $E$, magnetic field $B$, radius $R$ and energy gap $\Delta$. 
	
	\section{Elastic scattering}\label{SSPP}
	
	 {Before identifying the key parameters that characterize the scattering problem}, we first explain how an electron scatters on a circular GQD of radius $R$ in the presence of a  magnetic field. {Indeed}, consider an electron moving along the $x$-direction with an incident energy $E=\hbar v _F k$, where $k$ is the module of the wave vector $\vec k$. Then, the incident electron can be described by a plane wave as follows
	\begin{equation}\label{e:25}
		\Phi_k^i(r,\theta) =\frac{1}{\sqrt{2}}e^{ikr\cos\theta }\dbinom{1}{1}=\frac{1}{\sqrt{2}}\sum_{m=-\infty}^{\infty}i^m \dbinom{J_m (kr) e^{im \theta}}{iJ_{m+1}(kr)e^{i(m+1)\theta}}
	\end{equation}
	where $J_m(z)$ is the  Bessel function of the first kind. 
	The following equation demonstrates how the reflected electron wave can be divided into partial waves since it must adhere to infinite boundary requirements for the scattering mechanism being researched
	\cite{cserti2007caustics}	
	\begin{equation}\label{e:26}
		\Phi_k^r(r,\theta) =\frac{1}{\sqrt{2}}\sum_{m=-\infty}^{\infty} a_m^r i^m \dbinom{H_m (kr) e^{im\theta}}{iH_{m+1}(kr)e^{i(m+1)\theta}}
	\end{equation}
with $H_m(x)$ are the  Hankel functions of the first kind that are  linear combinations of $J_m$ and Neumann $Y_m$, i.e.,  $H_m(x) = J_m(x) + i Y_m(x)$. For a large value of $x$, their asymptotic behavior	are
	\begin{equation}\label{e:27}
		H_m(x) \underset{x\gg 1}{\sim} \sqrt{\frac{2}{\pi x}} e^{i(x-\frac{m\pi}{2}-\frac{\pi}{4})}.
	\end{equation}	
	The transmitted solution can be obtained from the previous analysis as
	\begin{equation}\label{e:28}
		\Phi_q^t(r,\theta) =\sum_{m=-\infty}^{-1} a_m^{t-}  \dbinom{\Phi_q^{A-} (r) e^{im\theta}}{i \Phi_q^{B-}(r)e^{i(m+1)\theta}} +\sum_{m=0}^{\infty} a_m^{t+ } \dbinom{\Phi_q^{A+} (r) e^{im\theta}}{i\Phi_q^{B+}(r)e^{i(m+1)\theta}}
	\end{equation}
	where $q$ represents the wave number associated to the electron inside the GQD, as shown in  Fig. \ref{figsystem}.

	To {study} the  scattering problem, we first calculate the scattering coefficients $a_m^r$ and $a_m^t$ using the continuity of eigenspinors at the boundary condition $r=R$. We stress that,  in contrast to the regular quadratic
		Hamiltonian, in this case the continuity of the wavefunction derivative is not required since the physical quantities (such as density current) do not involve any derivative. This process yields
	\begin{equation}\label{e:29}
		\Phi_k^i(R,\theta) +\Phi_k^r(R,\theta) =\Phi_q^t(R,\theta) 
	\end{equation}
giving rise to two conditions {of $a_m^r$ and $a_m^t$ represented by}
	\begin{subequations}\label{30}
		\begin{align}
			&	\frac{1}{\sqrt{2}} i^m J_m(kR)+\frac{1}{\sqrt{2}} i^m a_m^r H_m(kR) =a_m^t \Phi_q^{A\pm} (qR) \label{30a}\\
			&	\frac{1}{\sqrt{2}} i^{m+1} J_{m+1}(kR)+\frac{1}{\sqrt{2}} i^{m+1} a_m^r H_{m+1}(kR) = i a_m^t \Phi_q^{B\pm} (q R)  \label{30b} .
		\end{align}
	\end{subequations}
Consequently, we obtain 
	\begin{subequations}\label{31}
		\begin{align}
			a_m^{t\pm}&=\frac{i \sqrt{2} e^{im\pi/2}}{\pi q R[H_m(kR)\Phi_q^{B\pm} (qR)-H_{m+1}(kR)\Phi_q^{A\pm} (qR) ]} \label{31a}\\
			a_m^{r\pm}&=\frac{-J_m(kR)\Phi_q^{B\pm} (qR)+J_{m
					+1}(kR)\Phi_q^{A\pm} (qR)}{H_m(kR)\Phi_q^{B\pm} (qR)-H_{m
					+1}(kR)\Phi_q^{A\pm} (qR) }  \label{31b} .
		\end{align}
	\end{subequations}
	

	We now define the probability density function $\rho= \Phi^\dag \Phi$ and current density $j= \Phi^\dag \sigma \Phi$ using Hamiltonian  \eqref{ham2}, with the spinor $\Phi$ being dependent on the region where {$\Phi=\Phi_q^t$ \eqref{e:28} is inside the GQD and $\Phi=\Phi_k^i+\Phi_k^r$ (\ref{e:25}-\ref{e:26})} is outside. As a result, the radial component of {the reflected current associated with \eqref{e:26}} is given by
	\begin{equation}
		j_{\text{rad}}^r(\theta)={ (\Phi^r_k)^{\dagger}}\begin{pmatrix}			0 & \cos \theta-i \sin \theta \\
			\cos \theta+i \sin \theta & 0
		\end{pmatrix} {\Phi^r_k}.
	\end{equation}
	Taking the asymptotic behavior \eqref{e:27} into account, we calculate $ j_{\text{rad}}^r(\theta) $ as	
	\begin{equation}\label{e:32}
		j_{\text{rad}}^r(\theta) =\frac{4}{\pi k R}\sum_{m=-\infty}^{+\infty} \vert a_m^r\vert^2 +\frac{8}{\pi k R}\sum_{m<m'} \Re(a_m^r a_{m'}^r)\cos[(m-m')\theta].
	\end{equation}
In the limit  $k r \to \infty$, \eqref{e:32} is used to calculate the effective scattering cross section $\sigma$ defined by
	\begin{equation}
		\sigma=\frac{I_{\text{rad}}^r}{I^{\text{inc}} / A_u}
	\end{equation}
	where $I_{\text{rad}}^r$ represents the total reflected flux through the GQD of radius $R$ and {$I^{\text{inc}} / A_u$ denotes the incident flux per unit area}. Our calculation shows that $I_{\text{rad}}^r$ takes the form
	\begin{equation}
		I_{\text{rad}}^r=\int_0^{2 \pi} j_{\text{rad}}^r(\theta) r d \theta=\frac{8}{k} \sum_{m=-\infty}^{+\infty}\left|a_m^r\right|^2
	\end{equation}
and  $I^i / A_u=1$ for 
 the incident wave  \eqref{e:25}.
	
	To improve our study of the scattering problem of Dirac fermions in different sizes of circular quantum dots, we analyze the scattering efficiency $Q$. This is defined as the ratio of the division of the scattering cross section to the geometrical cross section  \cite{schulz2014electron}
	\begin{equation}\label{e:33}
		Q=\frac{\sigma}{2 R}=\frac{4}{k R}\sum_{m=-\infty}^{+\infty} \vert a_m^r\vert^2 .
	\end{equation}
{Note that by switching off the energy gap ($\Delta=0$ meV), we recover the results derived in \cite{pena2022electron}. This, in fact, shows that our results are general and will offer the possibility of discovering interesting properties of the present system.}

	\section{Numerical analysis}\label{res}

	We numerically examine the scattering phenomenon of electrons on a gapped GQD when it is exposed to a constant magnetic field. Our studies are based on the analysis of the magnitude in terms of scattering efficiency $Q$ and density $\rho$ in the region close to the quantum dot and the lifetime of the quasi-bound states.
	
	\subsection{Scattering efficiency}

\begin{figure}[ht]
	\centering
	\includegraphics[scale=0.33]{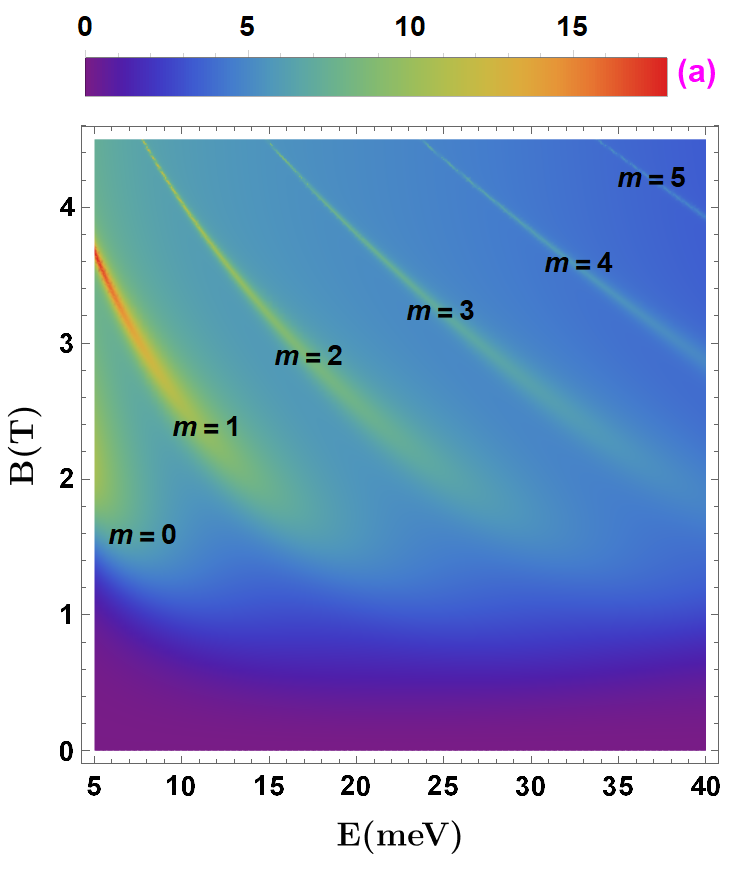}\includegraphics[scale=0.33]{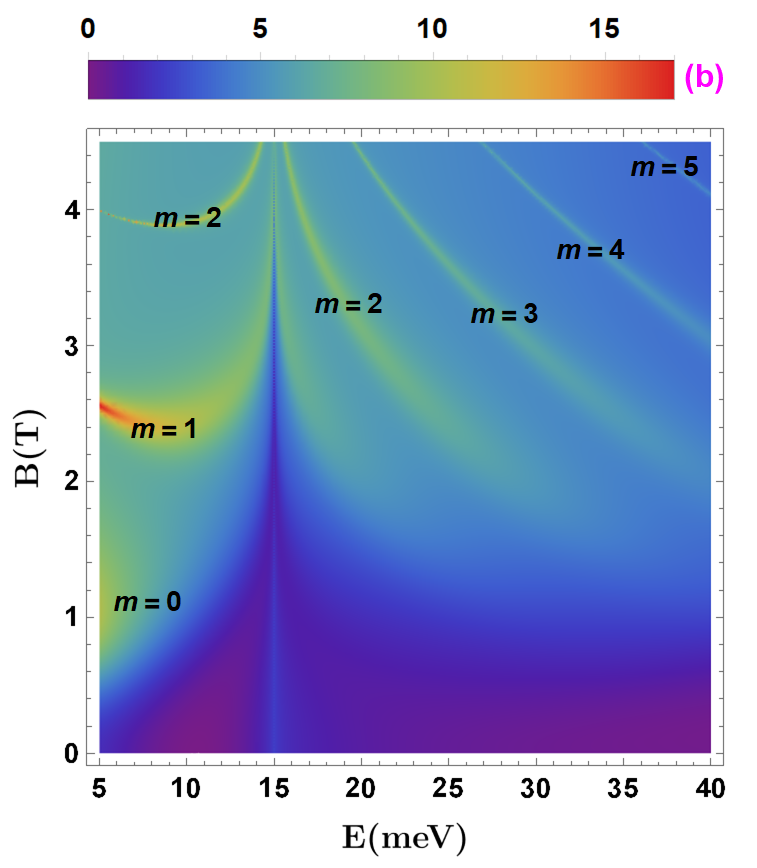}
	\includegraphics[scale=0.33]{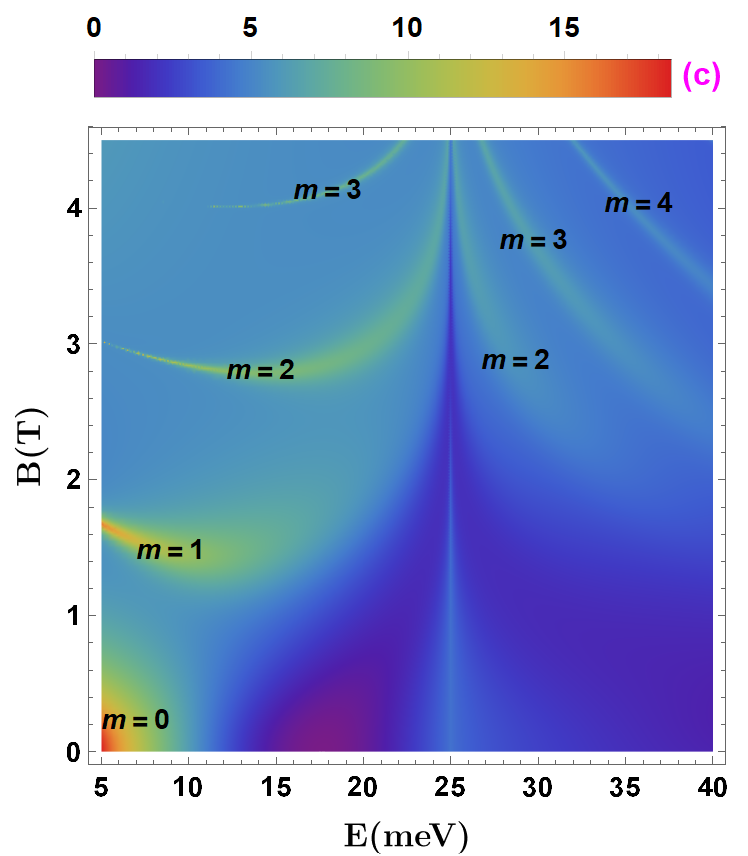}\includegraphics[scale=0.33]{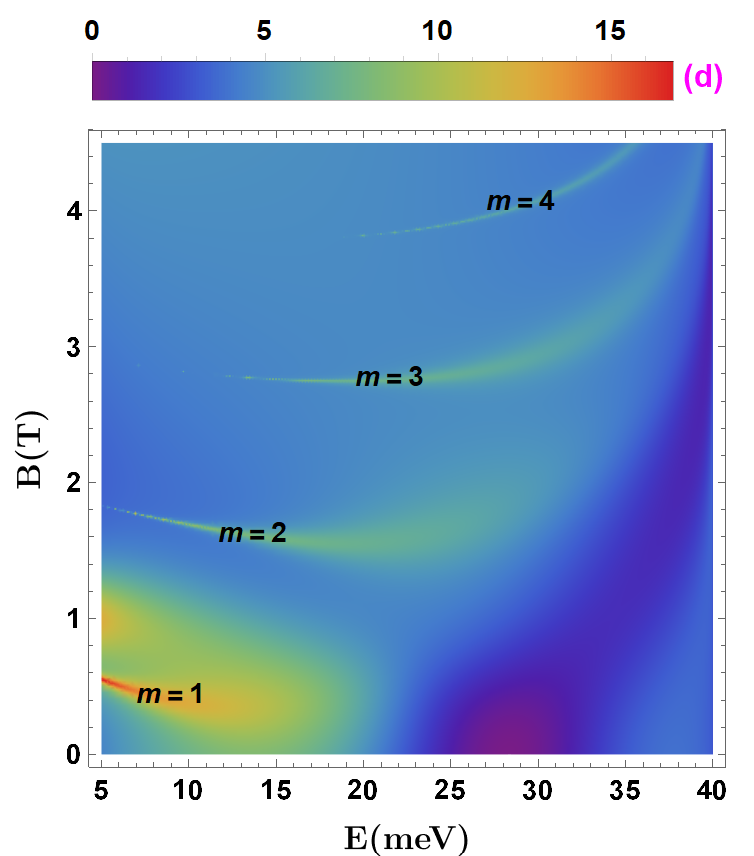}
	\caption{(color online) The scattering efficiency $Q$ as a function of the incident energy $E$ and  magnetic field  $B$ for $R=50$ nm different values of $\Delta$. (a): $\Delta=0$, (b): $\Delta=15$ meV, (c): $\Delta=25$ meV and (d): $\Delta=39.5$ meV}
	\label{fig2art}
\end{figure}

 {Fig. \ref{fig2art} displays a contour plot depicting the scattering efficiency, denoted as $Q$, as a function of incident energy $E$ and magnetic field strength $B$, while keeping the radius $R$ fixed at 50 nm. The plot considers various values of the energy gap $\Delta$ under different scenarios: In Fig. \ref{fig2art}a, where $\Delta$ equals 0 meV, the $Q$ pattern exhibits nearly oscillatory behavior, featuring six distinct bands characterized by high $Q$ values. Each of these bands corresponds to a scattering mode labeled as $m = 0, \cdots, 5$. This outcome aligns with previously reported findings in the published papers \cite{pena2022electron, chen2007fock, hewageegana2008electron, fu2020coulomb}. Furthermore, it is worth noting that the interaction strength remains notably weak until the magnetic field strength reaches approximately 1 T. As illustrated in Figs. \ref{fig2art}b, \ref{fig2art}c, and \ref{fig2art}d, when the energy gap $\Delta$ is introduced, it diminishes the resonance effect. Consequently, the interaction inside the Quantum Dot (GQD) commences at smaller radius values compared to the scenario in Fig. \ref{fig2art}a ($\Delta$ = 0 meV). This leads to an observable reduction in $Q$ as $\Delta$ gradually approaches the incident energy. Notably, with an increase in $\Delta$, the number of excited scattering modes gradually decreases, ultimately reducing to only four modes when $\Delta$ nears 40 meV. Another significant observation is that the interaction inside the GQD initiates even in the absence of a magnetic field ($B =$ 0 T), as evident in Figs. \ref{fig2art}c and \ref{fig2art}d.}

	\begin{figure}[H]
	\centering
	\includegraphics[scale=0.48]{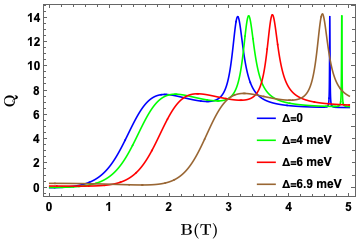}\includegraphics[scale=0.47]{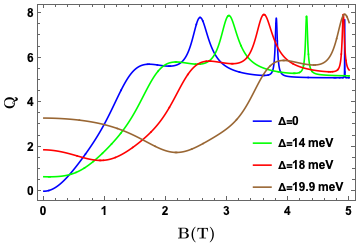}
	\includegraphics[scale=0.48]{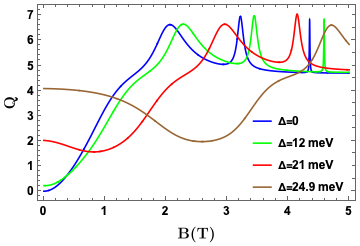}\includegraphics[scale=0.48]{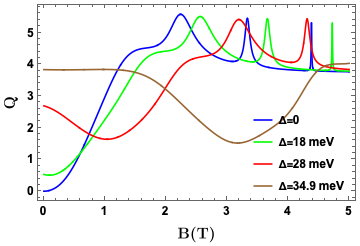}
	\caption{(color online) The scattering efficiency $Q$ as a function of the magnetic field  $ B $ for different values of  $\Delta$ and $E$. (a): $E=7$ meV and $\Delta = 0,4,6,6.9$ meV, (b): $E=20$ meV and $\Delta = 0,14,18,19.9$ meV, (c): $E=25$ meV and $\Delta = 0,12,21,24.9$ meV,  (d): $E=35$ meV and $\Delta = 0,18,28,34.9$ meV}
	\label{fig3art}
\end{figure}

The scattering efficiency $Q$ is plotted as a function of the magnetic field $B$ for various values of the  energy gap $\Delta$ and incident energy $E$ in Fig. \ref{fig3art}. 	
When $\Delta$ exceeds 6 meV for $E = 7$ meV and $B = 0$, $Q$ is not null, as shown in Fig. \ref{fig3art}a. Furthermore, we observe that $\Delta$ shifts the resonance peaks to the right, {which} leads to the suppression of a resonance peak at $\Delta \ge 6$ meV.
In Figs. \ref{fig3art}(b,c,d) {for} $E = 20 $, $25 $ and $35 $ meV,  {we} see that the resonance peaks are always shifted to the right due to the {the presence of $\Delta$}. Some peaks are also suppressed as $\Delta$ approaches $E$ and $Q$ takes specific values for $B=0$. This tells us that $Q$ survives and takes important values even in the absence of the magnetic field, except that one should be placed in the zone where $\Delta$ is closest to $E>$ 20 meV.

	\begin{figure}[H]
		\centering
		\includegraphics[scale=0.5]{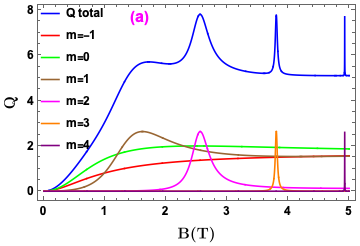}\includegraphics[scale=0.5]{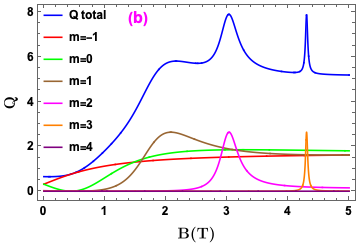}
		\includegraphics[scale=0.5]{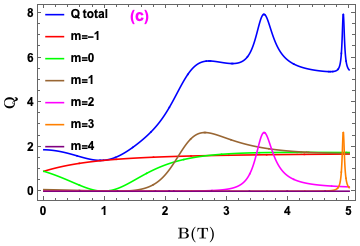}
		\caption{(color online) The scattering efficiency $Q$ as a function of the magnetic field $B$ for $E=20$ meV,  $m=-1,0,1,2,3,4$, and three values of the energy gap  (a): $\Delta=0$ meV, (b): $\Delta=14$ meV, (c): $\Delta=18$ meV.}
		\label{fig4art}
	\end{figure}

 We show the scattering efficiency $Q$ as a function of the magnetic field $B$ in Fig. \ref{fig4art} for E = 20 meV, three values of the energy gap $\Delta = 0$, 14,  18 meV and the quantum number $m = 0, \cdots, 5$. The first three modes, $m = -1, 0, 1$ are excited without resonance peaks  for $ \Delta = 0 $ meV, {as depicted  in Fig. \ref{fig4art}a}. On the other hand, {for  $m= 2, 3, 4$, we observe resonance peaks for specific values of $B$, whereas their width decreases with increasing $m$ and becomes extremely tiny at $m = 4$.} These results are in good agreement with those obtained in  \cite{pena2022electron}.	
 In Figs. \ref{fig4art}(b,c), we {notice} that the behavior of $Q$ is almost similar to that {seen} in  Fig. \ref{fig4art}a  except that the positions of the resonance peaks are shifted to the right. We also observe the suppression of the $m=4$ scattering mode. We emphasize  that in the absence of the magnetic field, the addition of $\Delta$ leads to a scattering efficiency  $Q$ no null, which becomes very important with the increase of  $\Delta$. 
 As shown in Figs.  \ref{fig4art}(b,c), $Q$ is solely due to the two scattering modes $m = -1, 0$. 
 
 	\begin{figure}[ht]
 	\centering
 	\includegraphics[scale=0.33]{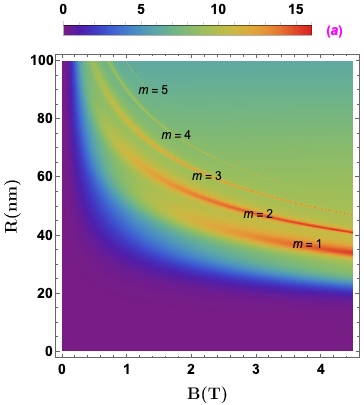}\includegraphics[scale=0.33]{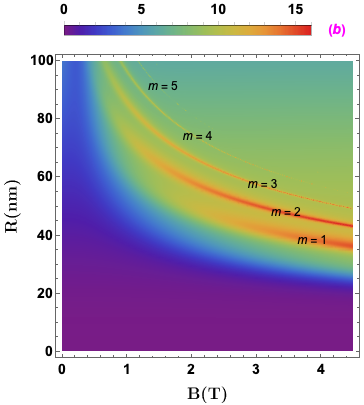}
 	\includegraphics[scale=0.33]{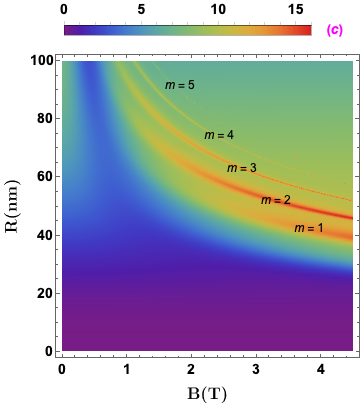}\includegraphics[scale=0.33]{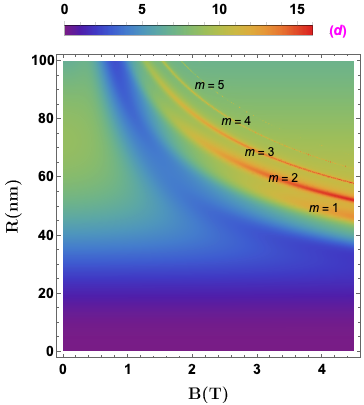}
 	\caption{(color online) The scattering efficiency $Q$ as a function
 		of the magnetic field $B$ and radius $R$ of the GQD for an incident energy value of $E= 20 $ meV and four values of the energy gap  (a): $\Delta=0$ meV, (b): $\Delta=14$ meV, (c): $\Delta=18$ meV, (d): $\Delta=19.9$ meV.}
 	\label{fig5art}
 \end{figure} 
 
 \begin{figure}[ht]
 	\centering
 	\includegraphics[scale=0.5]{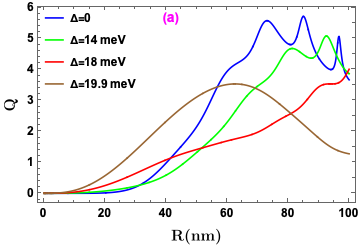}\includegraphics[scale=0.5]{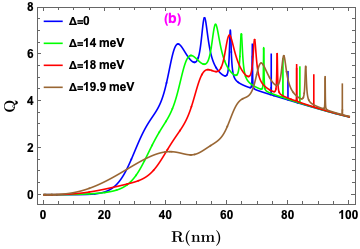}
 	\includegraphics[scale=0.5]{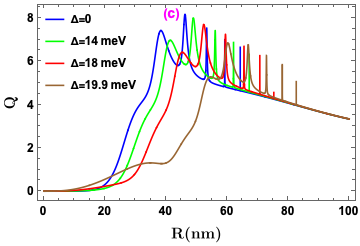}
 	\caption{(color online) The scattering efficiency $Q$ as a function of the radius $R$ of the GQD for three values of the magnetic field strength  (a): $B=0.8$ T, (b): $B=2.2$ T, (c): $B= 3.2$ T, and four values of the  energy gap $ \Delta=$ 0, 14, 18, 19.9 meV.}
 	\label{fig6art}
 \end{figure}
	
	The scattering efficiency $Q$ as a function of the magnetic field $B$ and radius $R$  of the GQD is shown in Fig. \ref{fig5art} for four different values of the energy gap {$\Delta=0,14,18,19.9$ meV}. In Fig. \ref{fig5art}a for $ \Delta=0 $, we observe that starting from $R \approx 32$ nm, the wide bands correspond to $m = 1, 2, 3$ and narrow bands correspond to $m = 4,5$ begin to show up.
	These important increases in $Q$ are designated as "scattering resonances" and are associated with a specific scattering mode. It is clear that the interaction is very low  below $R\approx 32$ nm.
	Furthermore, as shown in Fig. \ref{fig5art}a, the interaction is very weak inside the GQD, regardless of its radius, below $B \approx $0.4 T. 
%
Now by increasing  $\Delta$  in Figs. \ref{fig5art}(b,c,d), we see that {bands} appear for large values of $R$ and  grow as  $\Delta$ increases.  In particular, in Fig. \ref{fig5art}d,  the bands appear for $R=45$ nm and an energy gap $\Delta=19.9$ meV closes to the incident energy. This shows that the wide bands corresponding to the scattering modes ($m=1,\cdots,5$) shift to larger values of $R$ as long as $\Delta$ is increased. {It can be seen that the bands corresponding to the scattering modes (the interaction inside the quantum dot) start to become visible at a higher value of $R$ than in the case where $\Delta=0$.
The interaction is very weak inside the quantum dot, whatever its radius, below $B =0.4$ T and only from $R=0$ to $100$ nm. Now, if we increase $R$, the interaction starts to be visible for smaller values of $B$.
}

	Fig. \ref{fig6art} shows $Q$ as a function of $R$ for $E=20$ meV,   four values of $ \Delta$ and  magnetic field  (a): $ B=0.8 $ T, (b): $ B=2.2$ T,  (c): $ B=3.2$ T. 
	In Fig. \ref{fig6art}a, we observe that the most relevant resonance peaks are concentrated in the radius interval  $[50,100]$ nm. However,  this interval shifts to small values of $ R $  when  $B $ increases, as shown in Figs.  \ref{fig6art}(b,c). {This is related to the fact that the cyclotron radius varies inversely with the magnetic field.}
	More importantly, $Q$ becomes significant when $R$ is low and $\Delta$ is close to the incident energy. When $R$ is less than 20 nm, adding an energy gap $\Delta$ increases scattering efficiency $Q$, but when $R$ exceeds 20 nm, $Q$ decreases as $\Delta$ increases.

	\subsection{Probability density}
	\begin{figure}[H]
		\includegraphics[scale=0.343]{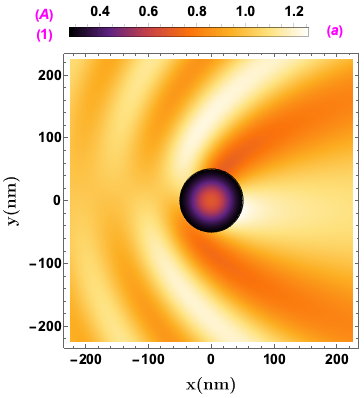}
		\includegraphics[scale=0.343]{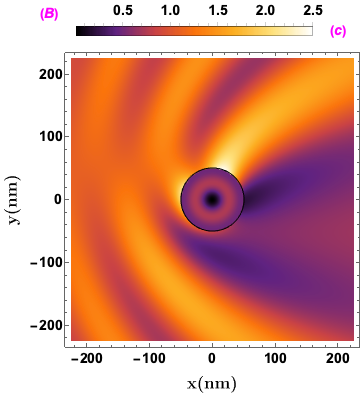}
		\includegraphics[scale=0.343]{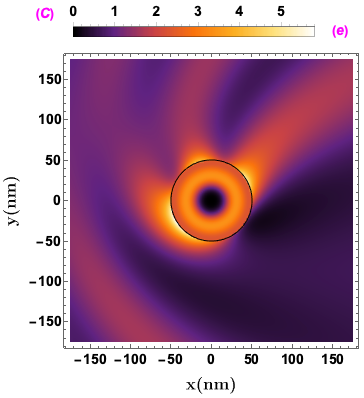}
		\includegraphics[scale=0.343]{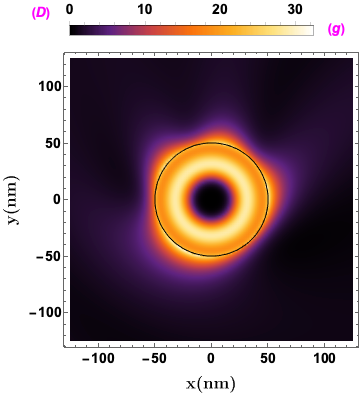}
		
		\includegraphics[scale=0.343]{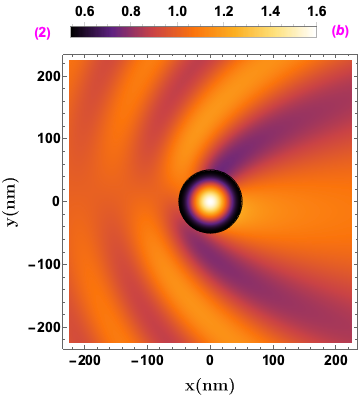}
		\includegraphics[scale=0.343]{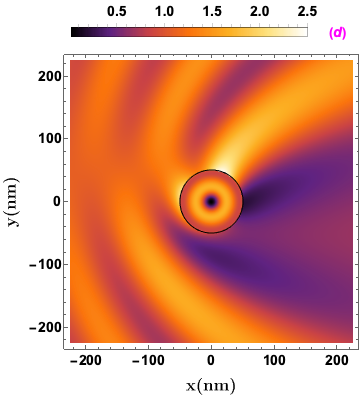}
		\includegraphics[scale=0.343]{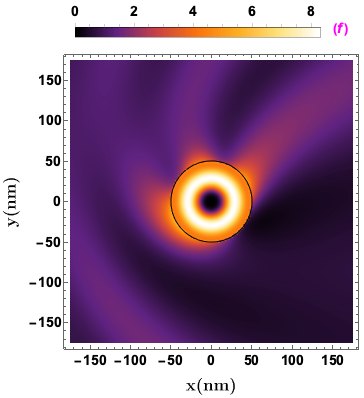}
		\includegraphics[scale=0.343]{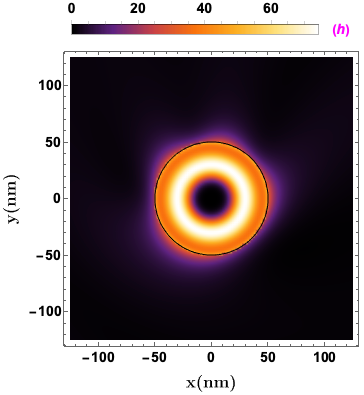}
		
		\caption{(color online) A density representation for a real space examination of electron scattering on a magnetically driven GQD. Each graph column is devoted to a given value of the quantum number  (A): $m=0$, (B): $m=1$, (C): $m=2$, and (D): $m=3$. Each graph line represents a specific value of the energy gap (1): $\Delta=0$ meV and (2): $\Delta=14$ meV. Each panel corresponds to a given value of magnetic field 
			(a): $B=0.4$ T, (b): $B=0.8$ T, (c): $B=1.62$ T, (d): $B=2.1$ T, (e): $B=2.56$ T, (f): $B=3.03$ T, (g): $B=3.8$ T, (h): $B=4.3$ T. A black circle indicates the geographic extent of the GQD.}
		\label{fig7art}
	\end{figure}
	
	In Fig. \ref{fig7art}, each column is dedicated to a scattering mode among those studied previously, and we examine the density in the field near the GQD for each mode under the effect of two values of the  energy gap $\Delta= 0$ and 14 meV. The boundary of the GQD is represented by the black circle. The magnetic field values chosen correspond to the peaks of  $m = 0, 1, 2, 3 $ presented in  Fig. \ref{fig4art}.
	Line \textcolor{red}{(1)} starts with the results for the four scattering modes $m=0, 1, 2, 3$ at zero energy gap. In Fig.  \ref{fig7art}a where $B=0.4$ T, we  see that the large values of the density are distributed in the outer part of the GQD, i.e., the electron wave is diffracted on the GQD boundary, as can be observed in Fig. \ref{fig4art}a where only the $m=-1,0$ modes are non-resonantly excited with a very low value of the scattering efficiency. Thus, in this case, we do not expect electron trapping effects inside the GQD. The mode $m = 1$ is resonantly excited with a broad peak in Fig. \ref{fig7art}c for $B = 1.62$ T. With the presence of the diffraction bands, we  also observe that the majority of the density is localized outside the GQD, but now the near field density values are slightly larger. 
	The mode $m = 2$ is excited with a narrower resonance peak in Fig. \ref{fig7art}e for $B = 2.56$ T than in the previous case. In this case, the majority of the density is concentrated inside the GQD with a high scattering efficiency. As a result, the small resonance peak of the mode $m=2$ makes it more likely that the electron will be trapped inside the GQD.
	The resonance of the mode $m=3$ is very clear in Fig. \ref{fig7art}g for $B=3.8$ T, with a narrower peak than for $m=2$. We  see the density concentration inside the GQD with a higher scattering efficiency and the suppression of diffraction bands. Therefore, the electron trapping effect inside the quantum dot is notable.
	From the aforementioned, we infer that the trapping effect increases with the resonance peak's narrowness.
	In line \textcolor{red}{(2)}, we	represent the results obtained at an  energy gap $\Delta=14$ meV for the four scattering modes $m=0,1,2,3$ with (b): ($m=0$,$B=0.8$ T) and (d): ($m=1$,$B=2.1$ T). It is clearly seen that the density inside and at the GQD boundary  is improved compared to those seen before, with the presence of diffraction bands around the GQD. Comparing (f): ($m=2$,$B=3.03$ T) and (h): ($m=3$,$B=4.3$ T) with (e) and (g), we see that the density is more concentrated inside the GQD and has created a clearer cloud around the core of the GQD. Therefore, when an energy gap $\Delta$ is added, the impact of electron trapping inside the GQD is improved.

	\subsection{Trapping time} 
	
	{Since the quasi-bound states are inherently unstable, it is necessary to estimate their lifetime (trapping time) in order to deeply understand our system's behavior. This will be clarified by outlining the mathematical process based on defining complex energy and wave number. It is worth noting that a quasi-bound state is distinguished by positive energy within the continuum as opposed to a real bound state, in which the energy spectrum is discrete and has negative values. The study of the lifetime of the quasi-bound state, can  be approached in terms of the complex-valued energies of the incident electron, denoted as $E = E_r - iE_i$ \cite{PhysRevLett.83.4991},  where $E_r$ represents the resonance energy while $E_i$ defines the lifetime of the quasi-bound states, which is $\tau=\frac{\hbar}{E_i}$. As a result, the wave number also becomes complex, and then we have $k = k_r - ik_i$. Considering the linear energy dispersion $E = v_F \hbar k$ and incorporating these factors, we can also have another expression $\tau=\frac{1}{v_Fk_i}$ {of lifetime}.
	}

	We have previously indicated that the whole scattering process can be decomposed into several scattering sub-processes. We will now concentrate on the two scattering modes $m=0,3$ to see how the energy gap $\Delta$ affects the lifetime  $\tau$ of the quasi-bound states.  
	We use the continuity to find the complex energy of the incident electron  by matching the transmitted wave function with the reflected wave function on the boundary $r = R$  \cite{hewageegana2008electron}.
	Since the incident energy of the incoming electron is not affected by the magnetic field, we treat the following transcendental equation for   $q$ and $k$ 
	\begin{equation}\label{e:34}
		\frac{\Phi_q^{A\pm} (qR)}{\Phi_q^{B\pm} (qR)}=\frac{H_m(k R)}{H_{m+1}(k R)}.
	\end{equation}		
	{This can be numerically resolved for each established resonance, and therefore the solution allows us to determine a set of incident energies $(E_r, E_i)$ associated with the magnetic field $B$. As a result, the lifetime $\tau$ of each quasi-bound state can be found.}

	\begin{figure}[H]
		\centering
		\includegraphics[scale=0.54]{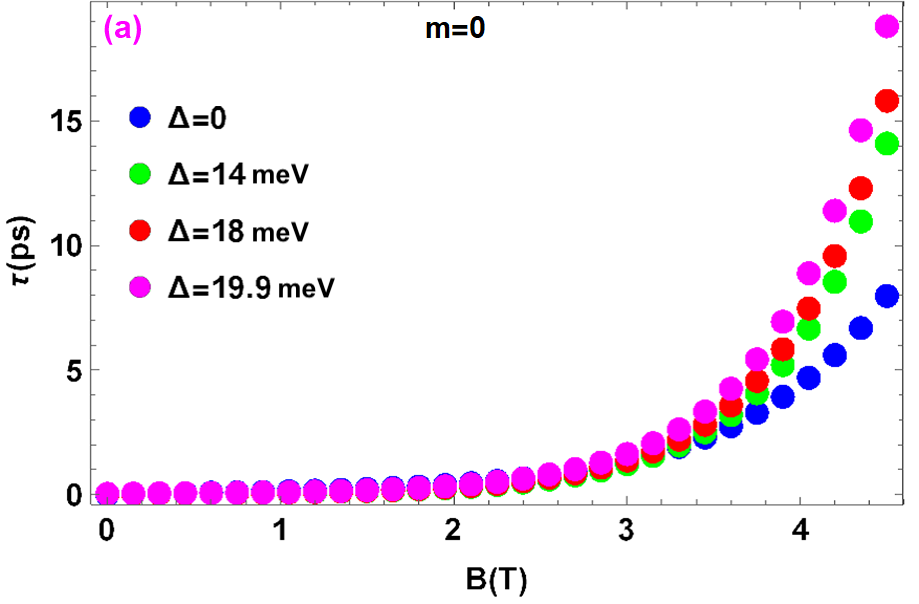}\includegraphics[scale=0.55]{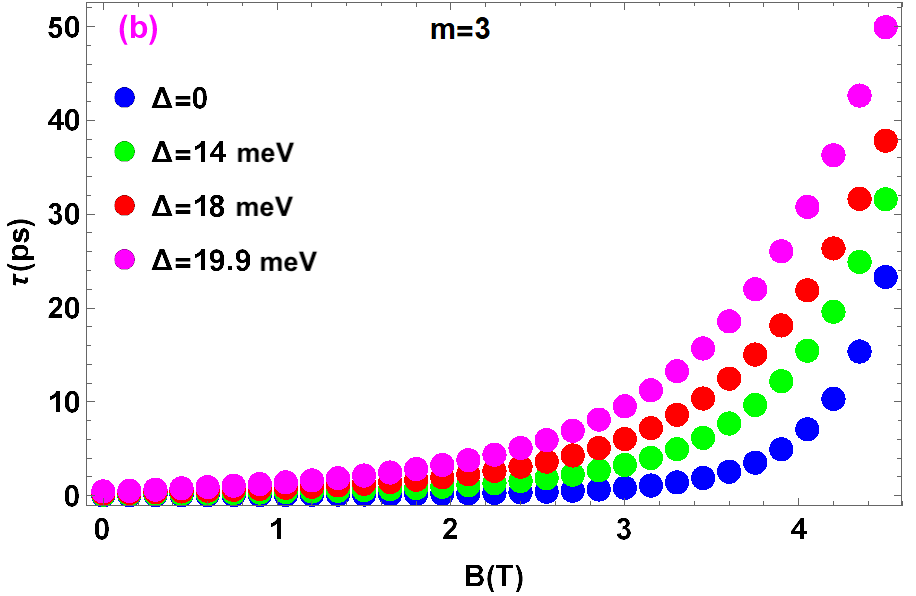}
		\caption{(color online) The lifetime $\tau$ as a function of magnetic field $B$ for  four values of the energy gap $\Delta=0,14,18,19.9$ meV and  two scattering modes (a): $m=0$, (b): $m=3$, {with $R=50$ nm.}}
		\label{fig8art}
	\end{figure}
	
	{To illustrate the above results, we choose two scattering modes $m=0,3$ and make a comparison between them. To this end, in Fig. \ref{fig8art}, we present the lifetime $\tau$ as a function of the magnetic field $B$ for} four values of the energy gap $\Delta$. The first thing we notice is that $ \tau $ increases as  $B$ increases, which agrees with what we previously discovered by examining the scattering efficiency $Q$.  
	In Fig. \ref{fig8art}a, $\tau$ {becomes appreciably different from zero} for $B=1.8$ T and increases more clearly from $B\approx 3.35$ T for $\Delta$  non-null. 
	In Fig. \ref{fig8art}b, we  see that for a given $\Delta$, $\tau$ tends to increase from small values of $ B $ when compared to the case of $\Delta=0 $ for  higher magnetic fields. We conclude that when $ \Delta $ is non-null, $\tau$ improves even further.

	\section{Conclusion} \label{cc}
	
	{We have theoretically studied the electron scattering mechanism in a graphene quantum dot (GQD), including a mass term, introducing an energy gap into the spectrum, and being subjected to a magnetic field. In this context, we initially established a theoretical framework for describing the interaction between Dirac fermions and a constant magnetic field within a circular GQD. To address this complex problem accurately, we began by solving the associated Dirac equation and obtaining analytical solutions for the energy spectrum. Employing the continuity conditions for the eigenspinors, we proceeded to compute various parameters that characterize the scattering phenomenon. Notably, one such parameter, the efficiency denoted as $Q$, was found to be dependent on several factors, including the magnetic field, angular momentum, GQD radius, incident energy, and energy gap.		
	}

We also presented numerically the obtained results to provide a general interpretation of the scattering phenomenon and to show how an electron at normal incidence can be trapped in a GQD for a certain period of time, with the main objective being to improve this trapping time. We discovered that even in the absence of a magnetic field, scattering efficiency can reach significant values when the energy gap is close to the incident energy of the electron crossing the GQD. For a non-null magnetic field, we showed that the resonance peaks with a higher scattering efficiency are those corresponding to the smallest values of  {the GQD radius}.

{Subsequently, we analyzed the probability density for two values of the energy gap $\Delta$. Indeed, for $\Delta=0$, we have shown that the diffraction phenomenon is dominant in the region where the scattering is non-resonant, with a weak localization of the density in the GQD. However, for $\Delta\neq 0$, it is found that the density inside the GQD is enhanced. On the other hand, in the region where the scattering is resonant, we found a damping of the diffraction with a very important localization of the density inside and at the boundary of the GQD. In addition, there are noticeable trapping effects.} The important result in our paper is to improve the possibility of trapping electrons in the GQD under the influence of a mass term, creating an energy gap in the energy spectrum.

	\begin{acknowledgments}
		We warmly thank Professor Adrian Pena for his valuable support.
	\end{acknowledgments}

	
\end{document}